\documentclass[aip,jap,reprint]{revtex4-1}

\usepackage{graphicx}% Include figure files
\usepackage{amsmath}
\usepackage{textcomp} %texterweiterungen
\usepackage{epstopdf}
\usepackage{epsfig}

\usepackage{color}

\begin{document}

\title{Excitation and detection of short-waved spin waves in ultrathin Ta/CoFeB/MgO-layer system suitable for spin-orbit-torque magnonics}

\author{T. Br\"{a}cher}
\email{thomas.braecher@cea.fr}
\affiliation{SPINTEC, UMR-8191, CEA-INAC/CNRS/UJF-Grenoble/Grenoble-INP, 17 Rue des Martyrs, 38054 Grenoble, France}
\author{M. Fabre}
\affiliation{SPINTEC, UMR-8191, CEA-INAC/CNRS/UJF-Grenoble/Grenoble-INP, 17 Rue des Martyrs, 38054 Grenoble, France}
\author{T. Meyer}
\affiliation{Fachbereich Physik and Landesforschungszentrum OPTIMAS, Technische Universit\"at Kaiserslautern, 67663 Kaiserslautern, Germany}
\author{T. Fischer}
\affiliation{Fachbereich Physik and Landesforschungszentrum OPTIMAS, Technische Universit\"at Kaiserslautern, 67663 Kaiserslautern, Germany}
\author{O. Boulle}
\affiliation{SPINTEC, UMR-8191, CEA-INAC/CNRS/UJF-Grenoble/Grenoble-INP, 17 Rue des Martyrs, 38054 Grenoble, France}
\author{U. Ebels}
\affiliation{SPINTEC, UMR-8191, CEA-INAC/CNRS/UJF-Grenoble/Grenoble-INP, 17 Rue des Martyrs, 38054 Grenoble, France}
\author{P. Pirro}
\affiliation{Fachbereich Physik and Landesforschungszentrum OPTIMAS, Technische Universit\"at Kaiserslautern, 67663 Kaiserslautern, Germany}
\author{G. Gaudin}
\affiliation{SPINTEC, UMR-8191, CEA-INAC/CNRS/UJF-Grenoble/Grenoble-INP, 17 Rue des Martyrs, 38054 Grenoble, France}

\date{\today}

\begin{abstract}

We report on the excitation and detection of short-waved spin waves with wave vectors up to about $40\,\mathrm{rad}\,\mu\mathrm{m}^{-1}$ in spin-wave waveguides made from ultrathin, in-plane magnetized Co$_{8}$Fe$_{72}$B$_{20}$ (CoFeB). The CoFeB is incorporated in a layer stack of Ta/CoFeB/Mgo, a layer system featuring large spin orbit torques and a large perpendicular magnetic anisotropy constant. The short-waved spin waves are excited by nanometric coplanar waveguides and are detected via spin rectification and microfocussed Brillouin light scattering spectroscopy. We show that the large perpendicular magnetic anisotropy benefits the spin-wave lifetime greatly, resulting in a lifetime comparable to bulk systems without interfacial damping. The presented results pave the way for the successful extension of magnonics to ultrathin asymmetric layer stacks featuring large spin orbit torques.
\end{abstract}


\maketitle

Magnonics and magnon spintronics\cite{Magnon-Spintronics,Neusser,Lenk-2011-1,Magnonics-Krug,Davies-2015-1,TAP-2013} explore the transport of information in the form of spin waves and magnons, their quanta, for post-CMOS devices as well as their interconnection with conventional electronics. The wave-nature of spin waves and their wavelength in the sub-micron range at microwave frequencies renders them an excellent candidate for a new generation of wave-based computing devices. The recently discovered exciting new ways to manipulate magnetization dynamics in ultrathin magnetic layers sandwiched in an asymmetric layer stack promise the creation of new magnonic devices and the replacement of Oersted fields by effective fields created by spin orbit torques (SOTs)\cite{Sampaio-2013-1,Manchon-2014-1,Miron-2010-1, Sklenar-2016-1}. These torques and the consequent fields can be created by a current flowing through the layer systems via spin-orbit coupling and the broken structural inversion symmetry. Unlike Oersted fields, they are localized to the current carrying interfaces. They allow to control the direction of the static magnetization\cite{Cubukcu-SOT,Fukami-SOT,BetaTa2}, the excitation of magnetization dynamics\cite{MockelPyPt, Demi-2014-STNO} and the control of the spin-wave damping in microscopic structures\cite{Gladii-2016-2}. However, most research\cite{MockelPyPt,Demi-2014-STNO,Gladii-2016-2} regarding spin waves and SOTs has been restricted to comparably thick films with thicknesses above $4-5\,\mathrm{nm}$. In contrast, SOTs are significantly more pronounced in ultrathin films with thicknesses well below $2\,\mathrm{nm}$ due to their interfacial origin\cite{Kim-2013}. In addition, many studies have been restricted to spin waves with wavelengths longer than $500\,\mathrm{nm}$, since these are comparably easy to excite and detect.

In this Letter, we report on the excitation and detection of short-waved spin waves with wavelengths down to about $\lambda \approx 150\,\mathrm{nm}$ in spin-wave conduits made from ultrathin Co$_8$Fe$_{72}$B$_{20}$ (CoFeB). The CoFeB features a thickness of about $d = 1.4\,\mathrm{nm}$ and is sandwiched in a Ta/CoFeB/MgO stack, a layer system which has been previously reported to feature large damping- and field-like SOTs\cite{CoFeB1,CoFeB2}. In addition, this layer system features a large perpendicular magnetic anisotropy (PMA) constant. At the investigated CoFeB thickness, the magnetization lies still within the film plane. Nevertheless, the large PMA contribution opposes the dipolar stray field and, this way, greatly reduces the dipolar effects on the spin-wave dispersion. Consequently, it lowers the spin-wave frequency and increases the spin-wave lifetime. We demonstrate that the spin-wave lifetime in the studied layer systems is comparable to the spin-wave lifetime in thick ferromagnetic films, despite the presence of strong interfacial damping in the ultrathin films. Consequently, these layer systems are an interesting alternative to thicker metallic layers with the additional benefit of large SOTs.

\begin{figure}[t]
	  \begin{center}
    \scalebox{1}{\includegraphics[width=0.9\linewidth, clip]{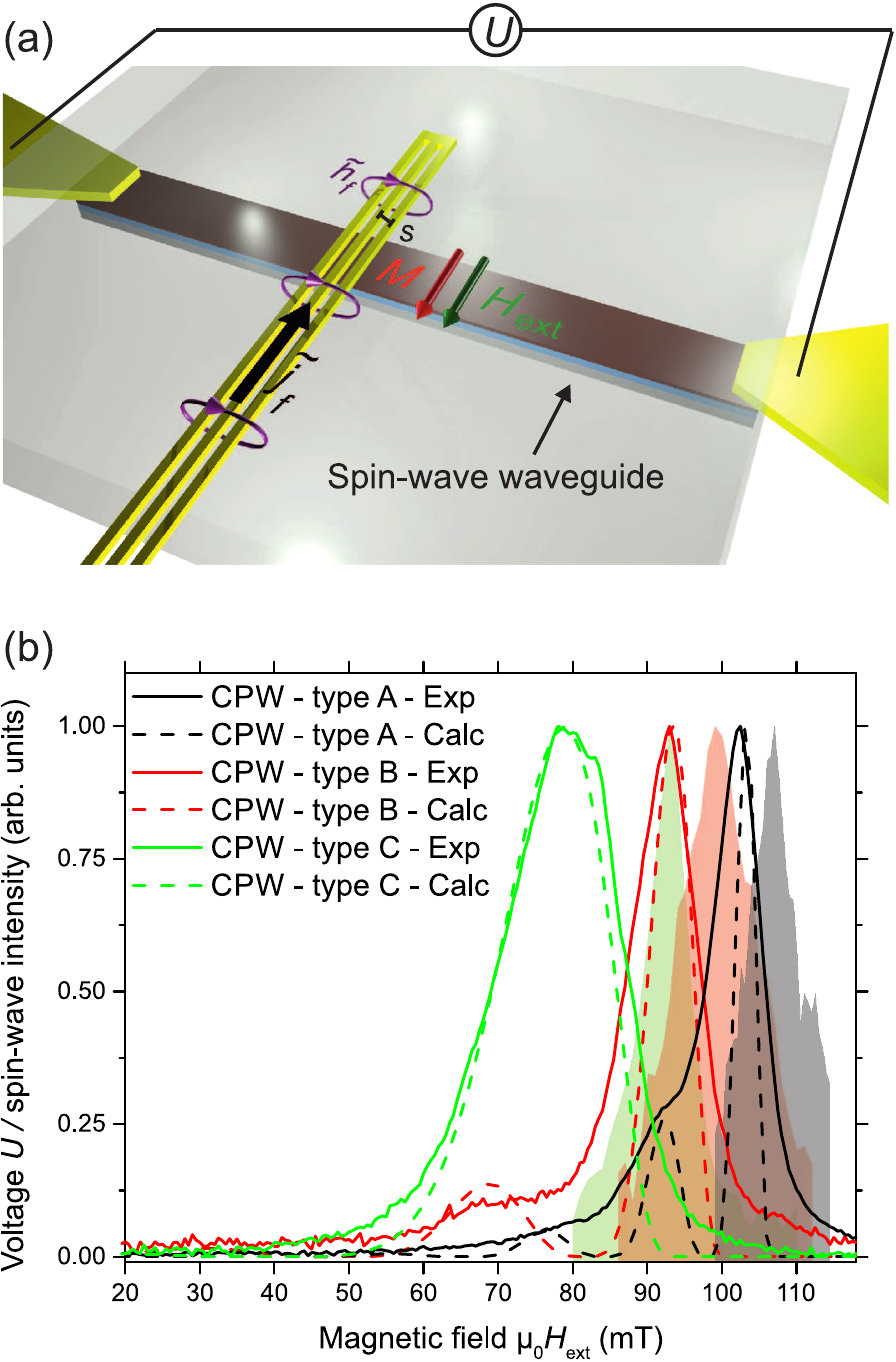}}
    \end{center}
	  \caption{\label{Fig1}(Color online) a) Schematic of the investigated sample and the geometry of the SWR measurements: A Ta/CoFeB/MgO trilayer is patterned into a spin-wave waveguide with leads to measure the voltage drop along the waveguide. On top of an insulating layer of Al$_2$O$_3$, a nanometric, shorted CPW with wire spacing $s$ acts as spin-wave excitation source. b) Exemplary excitation spectra at an excitation frequency of $4.8\,\mathrm{GHz}$ (solid lines) and analytical calculations of the expected excitation spectra of the CPWs (dashed lines). Black: CPW type A, red: CPW type B, green: CPW type C. The shaded peaks show exemplary STFMR spectra of the waveguides (see text for additional information).}
\end{figure}

The studied Ta/CoFeB/MgO trilayer has been deposited by sputter-deposition on an oxidized Si substrate. The CoFeB has been deposited in a wedge ranging from a thickness of $0.8\,\mathrm{nm}$ up to a thickness of $1.6\,\mathrm{nm}$ over the range of a 4~inch wafer and the trilayer is capped with an additional layer of Al$_2$O$_3$ and Ta$_2$O$_5$. The Ta thickness is $5\,\mathrm{nm}$, the thickness of the capping oxides is $1.5\,\mathrm{nm}$ (MgO), $2\,\mathrm{nm}$ (Al$_2$O$_3$) and $1\,\mathrm{nm}$ (Ta$_2$O$_5$), respectively. The used growth conditions ensure that the bottom Ta-layer grows in the $\beta$-phase, resulting in a large spin Hall angle within the Ta layer\cite{BetaTa,BetaTa2}. After an annealing step of the wafer at $250\,^\circ \mathrm{C}$ for $1.5\,\mathrm{h}$, the PMA is enhanced and leads to a transition to an out-of-plane magnetization of the CoFeB at thicknesses below approximately $1.2\,\mathrm{nm}$. Around a CoFeB thickness of about $1.4\,\mathrm{nm}$, which is still well in-plane, the trilayer is structured into spin-wave waveguides with a length of $10\,\mu\mathrm{m}$ and widths ranging from $500\,\mathrm{nm}$ to $5\,\mu\mathrm{m}$ by electron-beam lithography and ion-beam etching. Consequently, leads made from Ti/Au are defined at the short edges of the waveguides. These are used for the measurement of the rectified DC voltage created in the trilayer due to spin rectification if the magnetization is excited into precession\cite{STFMR1,STFMR2,STFMR3}. Subsequently, the waveguides are capped by a $30\,\mathrm{nm}$ thick layer of Al$_2$O$_3$ by means of atomic layer deposition. In a last series of steps, nanometric, shorted coplanar waveguides (CPWs) with different sizes are patterned on top of the waveguides by a combination of an electron beam-lithography and an optical lithograhpy step and electron beam evaporation. The CPWs are made from a double layer of Ti/Au with thicknesses of $5\,\mathrm{nm}$ and $30\,\mathrm{nm}$, respectively. They feature three different sizes and the corresponding structures are patterned in rows along the gradient in the CoFeB thickness. Type A features $100\,\mathrm{nm}$ wide wires with a $s = 500\,\mathrm{nm}$ center-to-center spacing, type B $50\,\mathrm{nm}$ wires with a $s = 300\,\mathrm{nm}$ center-to-center spacing and type C $50\,\mathrm{nm}$ wires with a $s = 150\,\mathrm{nm}$ center-to-center spacing. A schematic of the fabricated structures is shown in Fig. \ref{Fig1}~(a).

The reported electrical measurements have been performed on $5\,\mu\mathrm{m}$ wide waveguides, each featuring a different CPW of type A, B and C. For a first characterization of the properties of the waveguides, their dynamic properties are studied by means of spin torque ferromagnetic resonance spectroscopy (STFMR)\cite{STFMR1,STFMR2,STFMR3}. For these measurements, the CoFeB waveguides are positioned between a pair of coils under an angle of $45^\circ$ between the waveguides and the magnetic field within the film-plane. Using high-frequency probes, they are connected to a microwave generator via the capacitive part of a Bias-T. The generator provides an RF current leading to the excitation of magnetization dynamics in the waveguide as the magnetic field is swept across resonance. A lock-in amplifier (LIA) is connected to the inductive part of the Bias-T, which allows for a detection of the rectified DC voltage created by these magnetization dynamics within the waveguide. The LIA is used to modulate the amplitude of the microwave generator at a frequency of $10\,\mathrm{kHz}$ to enhance the signal-to-noise ratio in the experiment. 

\begin{figure*}[t!]
	  \begin{center}
    \scalebox{1}{\includegraphics[width=0.9\linewidth, clip]{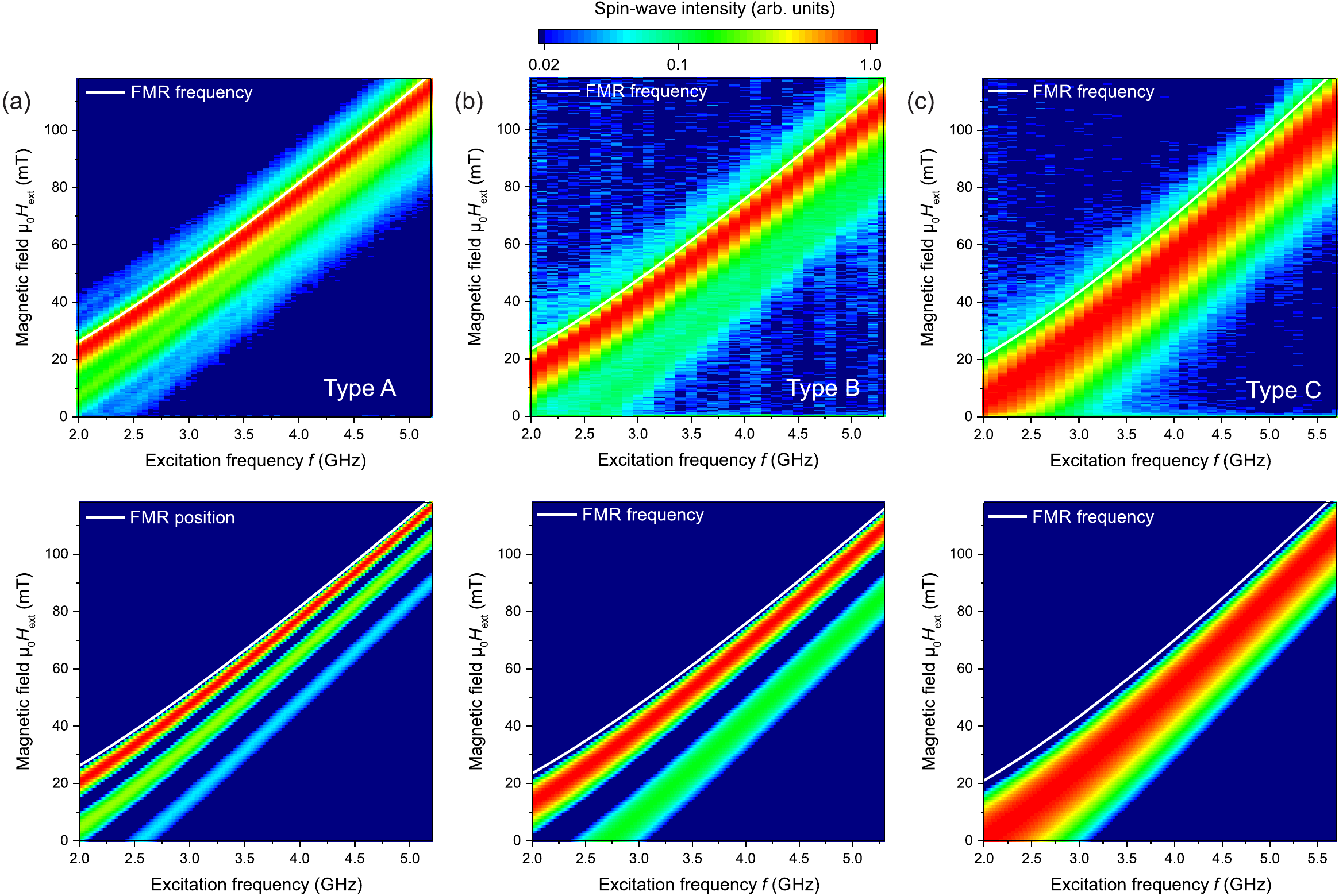}}
    \end{center}
	  \caption{\label{Fig2}(Color online) Color-coded measured SWR voltage/expected spin-wave intensities as a function of the applied frequency and applied magnetic field. The upper panel shows the measurement and the lower panel the analytical calculations. a) CPW type A: Wire width $100\,\mathrm{nm}$, wire spacing $s = 500\,\mathrm{nm}$. b) CPW type B: Wire width $50\,\mathrm{nm}$, wire spacing $s = 300\,\mathrm{nm}$. c) CPW type C: Wire width $50\,\mathrm{nm}$, wire spacing $s = 150\,\mathrm{nm}$. The white lines represent the Kittel fits obtained from the STFMR measurements. The calculations do not contain the broadening of the excitation peaks due to the finite damping in the material.}
\end{figure*}

Exemplary STFMR spectra are shown by the shaded peaks in Fig.~\ref{Fig1}~(b). From measurements of this kind, we determined the effective magnetization $M_\mathrm{eff} = M_\mathrm{s} - H_\perp$ in the waveguide by an analysis of the dependence of the resonance frequency on the applied magnetic field. Here, $M_\mathrm{s}$ denotes the saturation magnetization of the CoFeB and $H_\perp$ the anisotropy field due to the PMA. The resonance frequencies are fitted using the Kittel equation in the absence of any in-plane anisotropies, neglecting the small shape-anisotropy of the wire. This is justified due to the low thickness and the comparably large width of the wire. From this analysis, we find $M_\mathrm{eff,C} \approx 176\,\mathrm{kA}\,\mathrm{m}^{-1}$ for structures with CPWs of type C, $M_\mathrm{eff,B} \approx 154\,\mathrm{kA}\,\mathrm{m}^{-1}$ for type B and $M_\mathrm{eff,A} \approx 134\,\mathrm{kA}\,\mathrm{m}^{-1}$ for type A (the corresponding Kittel fits are shown by the white lines in Fig.~\ref{Fig2}). These different values are due to the different positions of the CPWs along the thickness gradient: From type C to type A, the thickness of the CoFeB reduces and, consequently, the anisotropy field $H_\perp$ slightly increases. Assuming a saturation magnetization\cite{These-Nistor} of $M_\mathrm{s} = 1250\,\mathrm{kA}\,\mathrm{m}^{-1}$, the obtained values of $M_\mathrm{eff}$ corresponds to an effective surface anisotropy constant of $K_\mathrm{\perp} \approx 0.60\,\mathrm{mJ}\,\mathrm{m}^{-2}$, a value in the expected range for Ta/CoFeB/MgO layer systems featuring the investigated CoFeB thickness\cite{These-Nistor,Ikeda}. In all waveguides we find field-linewidths on the order of a few $\mathrm{mT}$, featuring a linear dependence on the FMR-frequency. From the measured values, we extract an upper limit of the Gilbert-damping parameter of about $\alpha \approx 0.012$, assuming the entire damping is Gilbert-type.%\footnote{Since the frequency span we can cover in our STFMR measurements is rather low ranging from $2-5\,\mathrm{GHz}$ as a consequence of the low $M_\mathrm{eff}$ and the limited field-range of our coils, we cannot perform a trustworthy fit to obtain $\Delta H_0$ and $\alpha$ simultaneously.}. 

The relevant material parameters being determined, we now address the excitation of propagating spin waves with finite wave vectors in the waveguides. For an electrical characterization of the spin-wave excitation by the CPWs, these are connected to the microwave generator using high-frequency probes. The RF current provided by the generator creates a dynamic Oersted field around the CPWs. The leads at the waveguide edges are again connected to the LIA. In this kind of spin-wave rectification experiment (SWR), the rectified voltage arising from the spin-wave excitation in the spin-wave waveguide by the CPW is detected as a function of the applied field for a given excitation frequency. The waveguides are magnetized along their short-axis, to maximize the torque from the in-plane field created by the CPWs. In this geometry, the contribution of the inverse Spin Hall effect to the rectified voltage is maximal, while the contribution due to the anomalous magnetoresistance is minimum\cite{CoFeBSHNO}. Unlike STFMR, SWR gives access to spin waves with finite wave vectors, the excitation spectrum being defined by the geometry of the CPW.

The solid lines in Fig.~\ref{Fig1}~(b) show exemplary SWR measurements performed on the three different waveguides. The SWR voltage is proportional to the square of the dynamic magnetization and, thus, the spin-wave intensity. %Also in the SWR measurements, we find a purely symmetric peak shape. 
The measurements have been performed on the three different waveguides with different CPWs, using a microwave frequency of $4.8\,\mathrm{GHz}$ with an applied power of $P = 800\,\mu\mathrm{W} = -1\,\mathrm{dBm}$. This corresponds to the regime of linear excitation, whereas an increase of the power by about $2-3\,\mathrm{dB}$ leads to a deviation of the linear scaling of the measured voltage with the applied microwave power. The voltages have been normalized to their individual maximum, which is on the order of $2\,\mu\mathrm{V}$ for the applied power. As can be seen from the figure, the spacing between the FMR (shaded peak) and the peak of the CPW excitation gets larger and the width of the SWR peaks grows as the size of the CPW is reduced. This is due to the fact that the maximum spin-wave wave vector the CPW can excite increases. This is governed by the Fourier spectrum of the Oersted-field distribution created by the CPW\cite{Pirro-2011-1,Florin-2016}. To describe the spin-wave spectrum in the waveguides, we use the analytical formalism presented in Ref. \onlinecite{Braecher-2016-3} with the incorporation of the PMA discussed in the Supplemental material of this Letter. Assuming an effective width of the waveguide of $w_\mathrm{eff} = 5\,\mu\mathrm{m}$ and an exchange constant of $A_\mathrm{ex} = 10\,\mathrm{pJ}\,\mathrm{m}^{-1}$ as well as the other material parameters stated above, we calculate the expected spin-wave intensity spectrum excited by the CPWs which are represented by the dashed lines in Fig.~\ref{Fig1}~(b). Hereby, we average over the two emission directions along the wire, which are not equal due to the interplay of the in-plane and out-of-plane component of the Oersted field (cf., e.g., Ref. \onlinecite{Braecher-2016-3} and the comparison to the spectra obtained by Brillouin light scattering further down in this Letter). Furthermore, we only take into account the first waveguide mode in the calculations. The envelope of the analytical calculations is in good agreement with the experimentally obtained spectra. The main difference arises in the minima of the CPW excitation, which are situated at integer multiples of $2\pi\cdot s^{-1}$ and which cannot be completely resolved in the experiment due to the finite linewidth of the excited spin waves. This is, thus, mediated by the spin-wave damping, which is not included in the analytical formalism, and not due to the wave-vector dependent excitation properties of the CPWs. The envelope of the expected intensity drops below $2\,\%$ beyond the third minimum of the CPWs of type A, beyond the second minimum for type B and beyond the first minimum for type C. This wave vector corresponds to about $40\,\mathrm{rad}\,\mu\mathrm{m}^{-1}$ for all three types of CPWs and is equivalent to a wavelength of $\lambda = 166\,\mathrm{nm}$ for type A and of $\lambda = 150\,\mathrm{nm}$ for type B and C. For larger wave vectors, the relative intensity of any excited spin waves in the maximum of the excitation efficiency is below the noise of the used experimental setup. Up to this limit, %which is determined by the signal-to-noise ratio of the setup, 
the detection efficiency via the spin rectification seems independent of the spin-wave wave vector, since the measured spectra are solely determined by the features of the excitation source. This finding is in excellent agreement with a study of parametrically excited exchange magnons in macroscopic samples made from yttrium iron garnet\cite{Sandweg} and proves the versatility of this approach for the detection of traveling waves\cite{iSHE3}, even in micro- and nanostructures.

Figure~2 compares the measured excitation spectra of the three different CPW types to the corresponding expected excitation spectra. The measured voltage/expected spin-wave intensity is displayed color-coded as a function of the applied field and frequency. The spectra have been normalized individually to their maximum at each frequency to account for the changes of the input impedance of the CPW and of the used microwave equipment. All color maps use an identical, logarithmic scale. As can be seen from the figure, the measured spectra are in good qualitative and also quantitative agreement in the probed field- and frequency-range. In the entire range, the noise-limited maximum detectable wave-vector is about $40\,\mathrm{rad}\,\mu\mathrm{m}^{-1}$ and is determined by the Fourier spectrum of the excitation source. In addition to the color-coded spectra, the white lines correspond to the Kittel fits obtained from the STFMR measurements.

\begin{figure}[t!]
	  \begin{center}
    \scalebox{1}{\includegraphics[width=0.9\linewidth, clip]{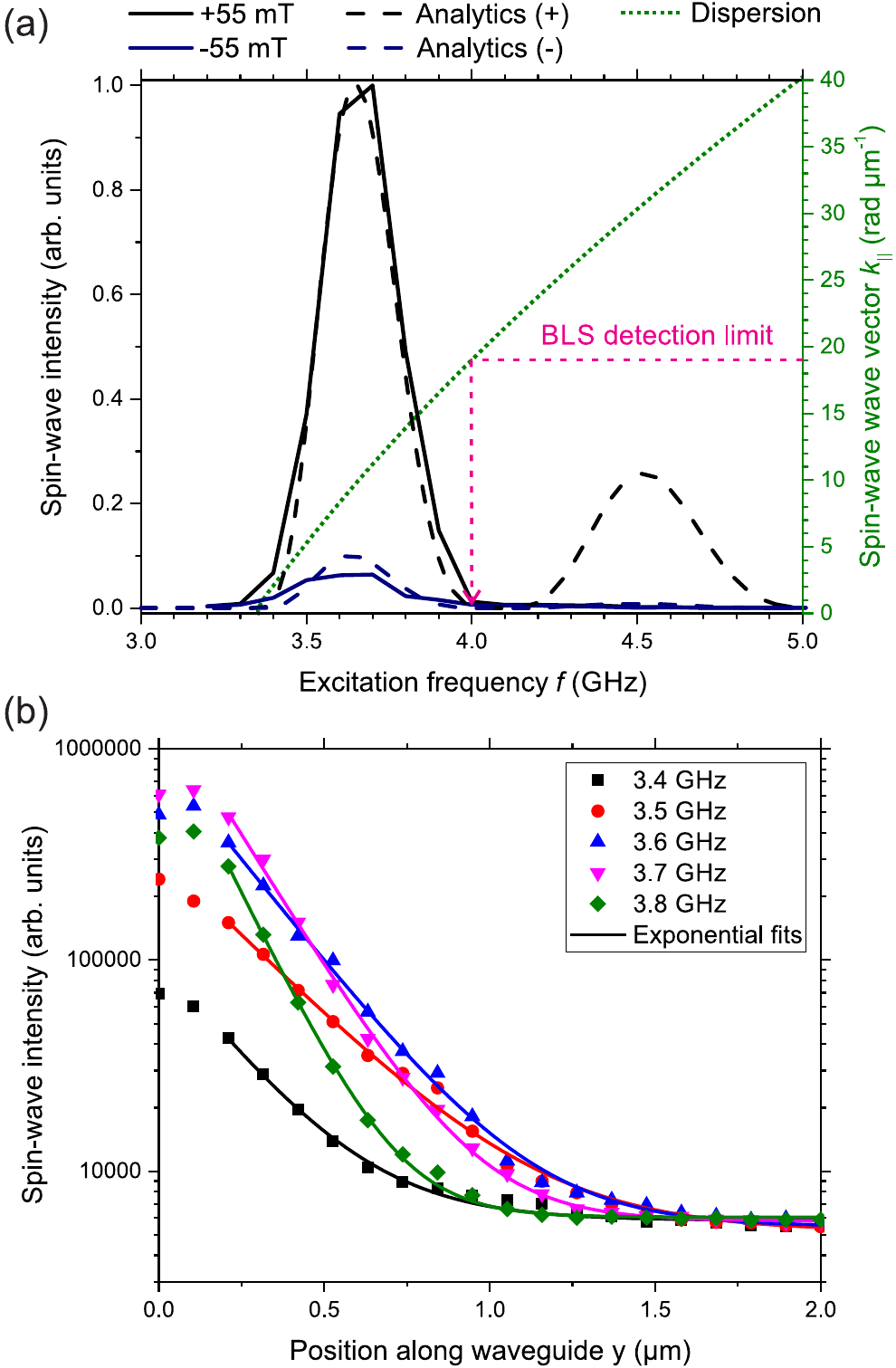}}
    \end{center}
	  \caption{\label{Fig3}(Color online) a) Measured (solid line) and calculated (dashed line) spin-wave spectra detected directly next to the antenna for $\pm \mu_0H\mathrm{ext} = 55\,\mathrm{mT}$. The dotted line shows the analytically calculated dispersion relation of the fundamental mode (right $y$-axis). The magenta line marks the BLS detection limit at $k \approx 19\,\mathrm{rad}\,\mu\mathrm{m}^{-1}$. b) Spin-wave intensity as a function of the position along the waveguide for different excitation frequencies. Solid lines are exponential fits.}
\end{figure}

To prove the propagating character of the excited waves, we study the excitation of a CPW of type B on a $2\,\mu\mathrm{m}$ wide waveguide by means of microfocused Brillouin light scattering (BLS)\cite{BLS2}. This technique gives access to the spin-wave dynamics for waves with wavelengths larger than approximately $300\,\mathrm{nm}$ with a spatial resolution of the same order of magnitude. In the BLS measurements, the field is fixed at $\mu_0|H_\mathrm{ext}| = 55\,\mathrm{mT}$ along the short axis of the waveguide. A microwave source is again connected to the CPW and the frequency is swept at an applied power of $1.26\,\mathrm{mW} = +1\,\mathrm{dBm}$. This larger power %is slightly above the threshold power for nonlinear excitation that has been observed in the electrical measurements but 
has been chosen to ensure a sufficiently large spin-wave excitation for a good signal-to-noise ration in the BLS experiment. Figure~\ref{Fig3}~(a) shows the BLS spectra measured directly next to the CPW for the two different field polarities. The measured spectrum in the vicinity of the CPW is compared to the analytically expected spin-wave excitation spectrum (dashed lines) and the spin-wave dispersion relation of the fundamental mode (dotted line). As can be seen from the figure, the two field polarities exhibit a strong asymmetry in terms of the overall intensity. Both, measurement and calculations, have been normalized to the maximum intensity in the efficient emission direction (in this case for positive magnetic fields). %As can be seen from the figure, the strong asymmetry can be nicely reproduced by the analytical calculation. In fact, t
The strong asymmetry is mediated by the PMA (see Supplementary material for more information): It decreases the ellipticity of precession and, this way, increases the influence of the out-of-plane component of the CPW fields on the spin-wave excitation. Consequently, its interference with the in-plane component becomes more pronounced. Thus, the use of CPWs or similar excitation sources for the spin-wave excitation in ultrathin films with large PMA results intrinsically in a pronounced uni-directional spin-wave emission. It should be noted that on the other side of the CPW, the opposite behavior is observed, i.e., the negative polarity is favored. By comparing the calculation (parameters of a structure of type B) to the experiment, it becomes evident that they only agree up to the first minimum of the CPW excitation. This is expected since this minimum, which corresponds to a spin-wave wavelength of $300\,\mathrm{nm}$, corresponds to the maximum wave vector which can be detected by the microfocussed BLS setup. Hence, for larger wave vectors, the BLS-detection becomes insensitive and, thus, the measured spin-wave spectrum does not reflect the wave-vector dependence of the excitation source anymore.

To analyze the spin-wave propagation within the waveguide, the spin-wave intensity is measured along the waveguide at different positions across its width. The intensity averaged across the width is shown as a function of the position along the waveguide for some exemplary frequencies within the detection range of BLS in Fig.~\ref{Fig3}~(b). Solid lines indicate exponential fits to the data. These have only been performed for distances which are sufficiently far from the CPW to ensure that the laser spot is not partially on the CPW. As can be seen from the fits, the spin-wave decay length initially increases with increasing frequencies as the decay becomes more shallow. However, for large frequencies, the decay becomes steeper. Around a frequency of $3.5\,\mathrm{GHz}$ it reaches its largest value of an amplitude decay length of about $600\,\mathrm{nm}$. This value is in excellent agreement with the prediction by the adopted analytical formalism presented in the Supplemental material, assuming the aforementioned values of $\alpha$ and $M_\mathrm{eff,B}$. It corresponds to a spin-wave amplitude lifetime of about $3\,\mathrm{ns}$ and a group velocity of about $200\,\mathrm{m}\,\mathrm{s}^{-1}$. The decrease of the decay length with increasing frequency is not expected from the analytical calculations, which predicts an approximately constant decay length. This indicates the presence of a wave-vector dependent damping mechanism in the measurements which is not incorporated into the analytical formalism. It should be noted that the lifetime of about $3\,\mathrm{ns}$ is large in comparison to the value expected from a material system featuring these magnetic parameters in the absence of PMA. This can be comprehended from the ellipticity contribution to the lifetime for the ferromagnetic resonance. The FMR lifetime\cite{MagDyn2} is given by $1/\tau = \alpha \gamma \mu_0 (H_\mathrm{ext} + M_\mathrm{eff}/2)$ and, thus, a reduction of $M_\mathrm{eff}$ significantly increases the lifetime at low magnetic fields. The obtained value of $3\,\mathrm{ns}$ is comparable to the lifetime in thicker ferromagnetic metallic materials such as Ni$_{81}$Fe$_{19}$ or the half-metallic Heusler compound CMFS in the absence of pronounced interfacial damping\cite{Bauer,Tomseb-2015}.

To conclude, we have demonstrated the excitation and detection of short-wave spin waves in spin-wave waveguides made from a Ta/CoFeB/MgO layer stack incorporating an ultrathin layer of CoFeB. We have demonstrated a wave-vector insensitive electrical detection by spin rectification up to a spin-wave wave vector of at least $40\,\mathrm{rad}\,\mu\mathrm{m}^{-1}$, which could be verified by comparing the measured voltage-spectra to the expected excitation spectra from the used nanometric CPWs. By employing microfocussed BLS, we have furthermore verified that the spin-wave emission by the CPWs exhibits a strongly preferred emission direction due to the large PMA in the investigated layer systems. The fact that this large PMA also results in a comparably large spin-wave lifetime and the value of the reported current-induced SOT in this material system renders it a highly interesting system for magnonic applications.

\begin{acknowledgements}
We thank St\'{e}phane Auffret et Marine Schott for their assistance in sample preparation. Financial support by the spOt project (318144) of the EC under the Seventh Framework Programme, by the DFG (SFB/TRR 173 Spin+X) and by the Nachwuchsring of the TU Kaiserslautern is gratefully acknowledged.
\end{acknowledgements}

\end{document}